\documentclass[11pt,a4paper]{article}
\usepackage{geometry}
\usepackage{graphics}
\usepackage{setspace}
\usepackage[scaled=0.9]{helvet}
\usepackage{graphicx}
\usepackage{subfig}
\usepackage{amssymb}
\usepackage{rotating}
\usepackage{arydshln}
\usepackage{abstract}
\usepackage[utf8x,utf8]{inputenc}
\usepackage[swedish, english]{babel}
\usepackage{fancyhdr}

\begin{document}
\bibliographystyle{acm}
\pagestyle{fancy}
\cfoot{\thepage}
\renewcommand{\abstractname}{}

\title{\fontfamily{phv}\selectfont{\huge{\bfseries{On the frequency distribution of neutral particles from low-energy 
strong interactions}}}}
\author{
{\fontfamily{ptm}\selectfont{\large{Federico Colecchia and Akram Khan}}}\thanks{Email: federico.colecchia@brunel.ac.uk}\\
{\fontfamily{ptm}\selectfont{\large{{\it Brunel University London, Kingston Lane, Uxbridge UB8 3PH, UNITED KINGDOM}}}}
}
\date{}
\maketitle
The rejection of the contamination, or background, from low-energy strong interactions at hadron collider experiments is a topic that has received significant attention in the field of particle physics. This article builds on a particle-level view of collision events, in line with recently-proposed subtraction methods. While conventional techniques in the field usually concentrate on probability distributions, our study is, to our knowledge, the first attempt at estimating the frequency distribution of background particles across the kinematic space inside individual collision events. In fact, while the probability distribution can generally be estimated given a model of low-energy strong interactions, the corresponding frequency distribution inside a single event typically deviates from the average and cannot be predicted a priori. We present preliminary results in this direction, and establish a connection between our technique and the particle weighting methods that have been the subject of recent investigation at the Large Hadron Collider.

~\\
{\bf Keywords:}
29.85.Fj; High Energy Physics; Particle Physics; Large Hadron Collider; LHC; soft QCD; pile-up; mixture models; Gibbs sampler; Markov Chain Monte Carlo; Expectation Maximisation.

\section{Nomenclature and general remarks}

\begin{itemize}
\item Collisions: proton-proton collisions at the Large Hadron Collider.
\item Events: triggered proton-proton collisions.
\item Bunch crossings: intersections between colliding proton beam bunches.
\item Physics processes: either the high-energy parton scattering of interest or low-energy strong interactions. 
\item Missing transverse energy: event-level energy imbalance measured on a plane perpendicular to the direction of the colliding particle beams.
\item Particle transverse momentum, $p_T$: absolute value of the component of the particle momentum vector on a plane perpendicular to the direction of the colliding beams.
\item Particle pseudorapidity, $\eta$: kinematic quantity expressed in terms of the particle polar angle in the laboratory frame, $\theta$, by $\eta=-\mbox{log}\left[\mbox{tan}(\theta/2)\right]$.
\item Whenever neutral particles are referred to in the text, neutrinos are not considered.
\end{itemize}

\section{Introduction}
The subtraction of the contamination, or background, from soft, i.e.\ low-energy, physics processes described by 
Quantum Chromodynamics (QCD) that take place in proton-proton collisions is a critical task at the Large Hadron Collider (LHC). 
The impact of the correction is going to become even more significant in the upcoming scenarios whereby the hard, 
i.e.\ high-energy, parton scattering of interest will be superimposed with a higher number of low-energy interactions from 
collisions between other protons, the so-called pile-up events. This is an important aspect at the LHC, and one that is going 
to have an even more significant impact at the High-Luminosity LHC (HL-LHC),  i.e.\ at the accelerator that is going to be 
built following the LHC upgrade project. In fact, the contribution of pile-up particles to the events of interest 
often makes the study of rare processes particularly challenging.

Subtraction techniques are well established, and typically combine tracking information for charged particles with 
estimates of the energy flow associated with neutral particles that originate from low-energy QCD interactions 
\cite{REVIEW_pile-up_2014}. In particular, pile-up subtraction is a key component in the data processing pipelines 
responsible for the reconstruction and calibration of jets, i.e.\ of collections of particles interpreted as originating 
from the same scattered parton. 

In this context, an important role has been played by correction procedures based on jet area \cite{jet-area}, 
which provides a measure of the susceptibility of reconstructed jets to the soft QCD energy flow. Such methods work 
by subtracting from the total momentum of the high-energy jets a quantity proportional to an event-level estimate 
of the background momentum density as well as to the area of the jet of interest. Therefore, this takes into account 
event-to-event background variability, and, since the correction can be calculated in a kinematics-depedent way, 
also the presence of different levels of pile-up activity in different kinematic regions inside events. 
However, due to the quantum nature of the underlying physics, the density of pile-up particles can be different even 
inside jets with very similar kinematics in the same collision event. While techniques based on jet area cannot correct 
for this, more recent methods exploiting information encoded in the substructure of jets 
\cite{jet-filtering,jet-trimming,jet-pruning-1,jet-pruning-2,jet-soft-drop,jet-cleansing} can effectively take this into account.

In this article, we explore a different perspective in order to estimate the frequency distribution of soft QCD particles 
inside events.
We build on a view of collision events as collections of particles whereby soft QCD particles are rejected 
upstream of jet reconstruction, in line with the particle-level pile-up subtraction algorithms that are currently being 
evaluated at the LHC \cite{PUPPI,SoftKiller,berta}.

Due to the quantum nature of the underlying physics processes and the limited number of final-state particles inside 
individual collisions, the particle multiplicity across the kinematic space inside events will generally vary across
collisions even when the physics processes involved are exactly the same. More precisely, 
the soft QCD particle-level frequency distribution will normally deviate from the corresponding probability distribution, 
and will be different in different events. What is discussed in this article is a data-driven method of estimating 
the soft QCD particle multiplicity across the kinematic space inside each event, using the following:

\begin{itemize}
\item The kinematic probability distributions of soft QCD particles and of particles originating from the signal hard scattering, 
e.g.\ obtained using simulated data;
\item The average fraction of soft QCD particles in the events;
\item The observed particle multiplicity, i.e. the observed number of particles in different kinematic regions in the event. 
\end{itemize}


Our approach relies on particles from high-energy scattering processes having pseudorapidity distribution more peaked at $\eta=0$, as well as higher values of $p_T$, on average, than those originating from low-energy strong interactions. This is essentially due to the higher transverse momentum transferred between the colliding protons, and results in different kinematic distributions of the final-state particles on the ($\eta, p_T$) plane, as illustrated in Fig. \ref{fig:cs} with reference to the control sample. Although different signal processes will generally be associated with different kinematic signatures, the dissimilarity between background soft QCD and hard scattering signal particles in terms of their $\eta$ and $p_T$ distributions typically outweighs the variability associated with the choice of signal process. 

Moreover, as a filtering stage upstream of physics analysis, reconstructed events at the experiments are usually 
subdivided, or ``skimmed'', into multiple data streams enriched in different signal processes. 
For the purpose of this discussion, the signal model can be thought of as describing the particle-level kinematics 
corresponding to the high-energy processes that the events analysed are enriched in. 

The kinematic variables used in this study are those that the relevant signal and background probability distributions 
can be written as functions of, namely particle pseudorapidity, $\eta$, and transverse momentum, $p_T$. 
To our knowledge, this is the first method of estimating how the 
frequency distribution of soft QCD particles 
inside individual events deviates from the expectation due to the non-deterministic nature of the underlying processes.

This article reports preliminary results on simulated collision events at the LHC, showing that the algorithm produces 
reasonable estimates of the number of soft QCD particles in different $(\eta, p_T)$ regions inside events regardless 
of the presence in those regions of particles from the hard scattering. Given that assessing the performance of 
this method requires knowledge of the true frequency distribution of soft QCD particles in each event, which is 
not available at the experiments, the validation was performed on simulated data, using an event generator commonly 
employed in the field \cite{pythia1,pythia2}. Specifically, background and signal particles were generated using 
soft QCD processes and $gg\rightarrow t\bar{t}$, respectively. 

Our interest in the estimation of the 
multiplicity of soft QCD particles across the kinematic space 
inside individual collision events 
relates to the development of further-improved methods of rejecting pile-up in high-luminosity regimes. 
Since our approach is based on a different principle and works in a different way as compared to established techniques, 
we expect its combined use with existing methods to result in enhanced pile-up subtraction in the upcoming 
higher-luminosity regimes at the LHC. We speculate that this can also lead to improved missing transverse energy resolution 
and to higher-quality estimates of particle isolation as the pile-up rates increase.

It should be noted that, in addition to pile-up, another source of soft QCD particles at the LHC is the so called 
Underlying Event (UE), which consists of particles from low-energy parton interactions taking place in the same 
proton-proton collision that contains the particles produced by the hard parton scattering. 
Pile-up and UE particles originate from similar processes: for this reason, with regard to estimating the 
frequency distribution of soft QCD particles inside events, it is expected that UE particles will contribute 
to the background category, i.e.\ that they will count towards the number of soft QCD particles together with 
those that originate from pile-up. In any case, although the distinction between pile-up and UE particles 
is conceptually important, pile-up is by far the primary source of soft QCD contamination at the ATLAS and CMS 
experiments at the LHC.

The algorithm that we describe in this article is a simplified deterministic variant of the Markov Chain Monte Carlo 
technique used in \cite{gibbshep3,gibbshep2,gibbshep}, where we discussed the idea of filtering collision events 
particle by particle upstream of jet reconstruction. Both our previous contributions and the present article 
relate to the development of new subtraction methods, with a view to improving further on the rejection of 
contamination from low-energy strong interactions in high-luminosity hadron collider environments. 
In particular, it is our opinion that the simplicity and parallelisation potential of this technique make 
it a promising candidate for inclusion in particle-by-particle event filtering procedures at the reconstruction 
level at future high-luminosity hadron collider experiments.

\section{The algorithm\label{algo}}

By construction, the probability density function (PDF) describing the kinematics of particles originating from a given process, 
e.g.\ with reference to soft QCD interactions, 
can be estimated as the limit of the corresponding frequency distribution, 
averaged over a large enough number of events. 
On the other hand, the corresponding frequency distribution inside a single event normally deviates from the PDF due 
to the limited number of particles. In fact, even when the processes involved are exactly the same, different 
collisions contain independent, and therefore different, realisations of the underlying quantum processes. 
For this reason, the shapes of the corresponding particle-level frequency distributions, 
e.g.\ that of soft QCD particles, generally vary across collisions. 

Let $f_0$ and $f_1$ denote the kinematic PDFs of background and signal particles, respectively, 
normalised such that $\int\int f_i(\eta, p_T) d\eta dp_T = 1,~i=0,1$. For the purpose of this study, we describe 
collision events as statistical populations of particles originating from soft QCD interactions and from the hard scattering, 
using a mixture model of the form $\alpha_0 f_0(\eta, p_T) + \alpha_1 f_1(\eta, p_T)$, where $\alpha_0$ is the fraction of 
soft QCD particles in the events, and $\alpha_1 = 1-\alpha_0$.

In this context, the probability for a given particle to originate from soft QCD interactions can be expressed using 
the following quantity:

\begin{equation}
w_0(\eta, p_T) \equiv \displaystyle \frac{\alpha_0 f_0(\eta, p_T)}{\alpha_0 f_0(\eta, p_T) + \alpha_1 f_1(\eta, p_T)}.
\label{eq:w0}
\end{equation}

Inside each collision event, although the actual numbers of background and signal particles in the different 
$(\eta, p_T)$ bins are not known, it is possible to estimate the corresponding expected numbers, 
$\nu_b(\eta, p_T)$ and $\nu_s(\eta, p_T)$, given a background and a signal model, respectively. 

For the purpose of this discussion, we express $\nu_b$ in terms of 
$\nu_b(\eta, p_T) = N \alpha_0 f_0(\eta, p_T)\Delta\eta \Delta p_T$, where $N$ is the total number of particles in the event, 
and $\Delta\eta$ and $\Delta p_T$ are the bin widths along the $\eta$ and $p_T$ axes, respectively. 
The corresponding expected number of signal particles in the bin, $\nu_s(\eta, p_T)$, can be calculated in a similar way 
using $f_1$.

If one denotes the unknown true numbers of signal and soft QCD particles as functions of particle 
$\eta$ and $p_T$ in each event by $n_s^*(\eta, p_T)$ and $n_b^*(\eta, p_T)$, respectively, then 
$n(\eta, p_T) = n_s^*(\eta, p_T) + n_b^*(\eta, p_T)$, where $n(\eta, p_T)$ is the corresponding 
number of particles in the data. When one considers LHC events with a number of proton-proton interactions in line with 
what is expected in the upcoming higher-luminosity regimes, the final-state particle multiplicities associated with 
soft QCD interactions and with the signal hard scattering are such that the expected number of signal particles 
in each bin, $\nu_s(\eta, p_T)$, is on average much lower than the corresponding number of soft QCD particles, 
i.e. $\left<\nu_s(\eta, p_T)\right> \ll \left<n_b^*(\eta, p_T)\right>$, where the average is taken over the $(\eta, p_T)$ space. 

One therefore expects that the statistical fluctuations on the observed number of particles in each $(\eta, p_T)$ bin 
will also be dominated by those on the number of soft QCD particles, 
i.e.\ $\left<\sigma_{n_s}(\eta, p_T)\right> \ll \left<\sigma_{n_b}(\eta, p_T)\right>$.

It should be noted that, if the variability of the number of signal particles can be neglected, the quantity

\begin{equation}
\hat{n}_b(\eta, p_T) = w_0(\eta, p_T) n(\eta, p_T)
\label{eq:nbhat}
\end{equation}

is expected to provide a more reliable estimate of the unknown number of soft QCD particles, 
$n^*_b(\eta, p_T)$, than $\nu_b(\eta, p_T)$ does. In fact, if the true number of soft QCD particles in each 
$(\eta, p_T)$ bin inside a given event deviates from the expected value by an amount $\Delta n_b$, 
i.e.\ if $n^*_b = \nu_b + \Delta n_b$, then a fraction $w_0$ of $\Delta n_b$ will contribute to $\hat{n}_b$ in that bin. 
On the other hand, $\nu_b$ reflects an expectation and therefore does not contain any information about statistical 
fluctuations inside events.

Given the estimated number of soft QCD particles in each ($\eta, p_T$) bin, $\hat{n}_b(\eta, p_T)$, the corresponding 
unknown actual number can be treated as a random variable following a binomial distribution with mean given by expression 
(\ref{eq:nbhat}) and standard deviation

\begin{equation}
\sigma_{n_b} = \sqrt{n w_0 (1-w_0)}.
\label{eq:sigma_nbhat}
\end{equation}

Based on expression (\ref{eq:w0}), for the purpose of estimating the 
background particle multiplicity inside each event, 
the discrimination between soft QCD interactions and the hard parton scattering exploits the different shapes 
of the corresponding PDFs as functions of $\eta$ and $p_T$. Specifically, as anticipated, the discrimination relies 
on particles from the hard scattering 
having a pseudorapidity distribution more peaked at $\eta=0$, 
as well as having on average higher values of $p_T$.

The use of expression (\ref{eq:w0}) for $w_0(\eta, p_T)$ in order to estimate the 
multiplicity of soft QCD particles across the kinematic space 
inside the events is essentially equivalent to weighting particles against PDFs that summarise our knowledge of the 
kinematics of the underlying processes, thereby taking into account the average fraction of soft QCD particles in the events. 
As far as the hard scattering is concerned, in addition to $gg\rightarrow t\bar{t}$, which is used to illustrate 
our method in the present article, the algorithm has also been run on particles originating from decays of the 
Standard Model Higgs boson produced via vector boson fusion \cite{CHEP2015_arxiv}. 
In fact, such a process does not involve colour exchange 
between the colliding protons, and is therefore expected to lead to a lower degree of particle activity around 
$\eta=0$ when compared to $gg\rightarrow t\bar{t}$.

The following discussion refers to neutral particles, since the identification of charged pile-up particles is made 
significantly easier by the availability of information from the tracking detectors at the experiments. 

It is worth pointing out that, 
although the signal and background PDFs were derived using simulated collision events in the context of this study, 
similar information can in principle be obtained using control samples from the data. As for the estimation of the 
soft QCD particle fraction among the neutral particles in the events, $\alpha_0$, it was decided to use the corresponding 
fraction of charged particles averaged over the events generated. In fact, the investigation of a more sophisticated 
approach including the use of a kinematic correction factor obtained from Monte Carlo showed no significant performance 
improvement \cite{fingerprints_arxiv}. It should also be emphasised that the present estimate of $\alpha_0$ based on 
charged particles was exclusively obtained for the purpose of this investigation, and that more information will typically 
be available at the experiments, e.g.\ in the form of data on the neutral energy flow provided by the calorimeters.

Whereas $\alpha_0$ was defined as a global event-level quantity, the possibility of introducing a dependence 
on $\eta$ and $p_T$ is worth investigating in the future, as this could lead to further-improved results. 
Nonetheless, it was decided to rely on as simple an approach as possible for the purpose of this feasibility study.

\begin{figure*}
\centering
\includegraphics[scale=0.4]{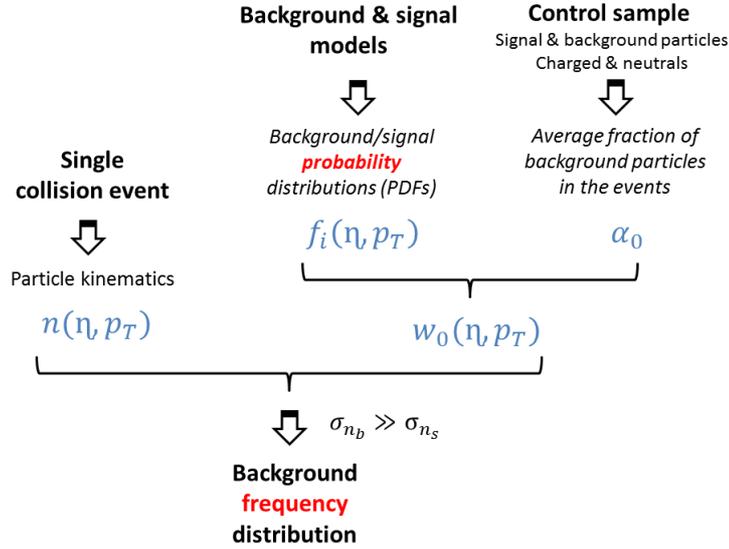}
\caption[]{
Overview of the method proposed in this article. Information from the background and signal models, as well as from the data, 
is combined into an estimate of the number of soft QCD particles across the particle ($\eta, p_T$) space 
inside each collision event, 
$\hat{n}_b(\eta, p_T)$, under the assumption that the statistical fluctuations in the data are dominated by the 
fluctuations on the number of soft QCD particles. The background and signal PDFs are denoted by $f_0$ and $f_1$, respectively.
}
\label{fig:overview}
\end{figure*}

An overview of this method is given in Fig. \ref{fig:overview}, which highlights what information is extracted from 
the models and what comes from the data. We will show that this approach produces a more accurate estimate of the 
background particle 
multiplicity 
inside the events than would be obtained using exclusively the expected number, $\nu_b$, as long as the statistical 
fluctuations in the data are dominated by those on the number of soft QCD particles. 

\section{Results\label{results}}

We discuss the results of a proof-of-concept study on Monte Carlo data at the generator level. 
We used Pythia 8.176 \cite{pythia1, pythia2} to generate 1,000 events, each consisting of a 
$gg\rightarrow t\bar{t}$ hard parton scattering at $\sqrt{s} =$ 14~TeV, superimposed with 50 soft QCD interactions to simulate the presence of pile-up. Soft QCD interactions were generated with ``SoftQCD:all'', ``PartonLevel:ISR'', ``PartonLevel:FSR'' and ``PartonLevel:MI'' set to ``on''. The performance of this method on a reference event will be discussed for the sake of illustration, and distributions on all events generated will then be shown in order to confirm the consistency of the results.

As a prerequisite for the execution of this method, the particle-level $(\eta, p_T)$ space in each event was subdivided 
into bins of widths $\Delta\eta = 0.5$ and $\Delta p_T = 0.05~\mbox{GeV}/c$. Whereas our $\Delta p_T$ binning will have 
to be revised in the context of a full detector simulation study, which is outside the scope of this article, we are using 
this choice of bins as a starting point to illustrate our method. Our analysis focusses on particles with 
$0 < p_T < 1~\mbox{GeV}/c$, which are the majority of those produced by soft QCD interactions.

\begin{figure*}
\centering
\subfloat[]{
\includegraphics[scale=0.3]{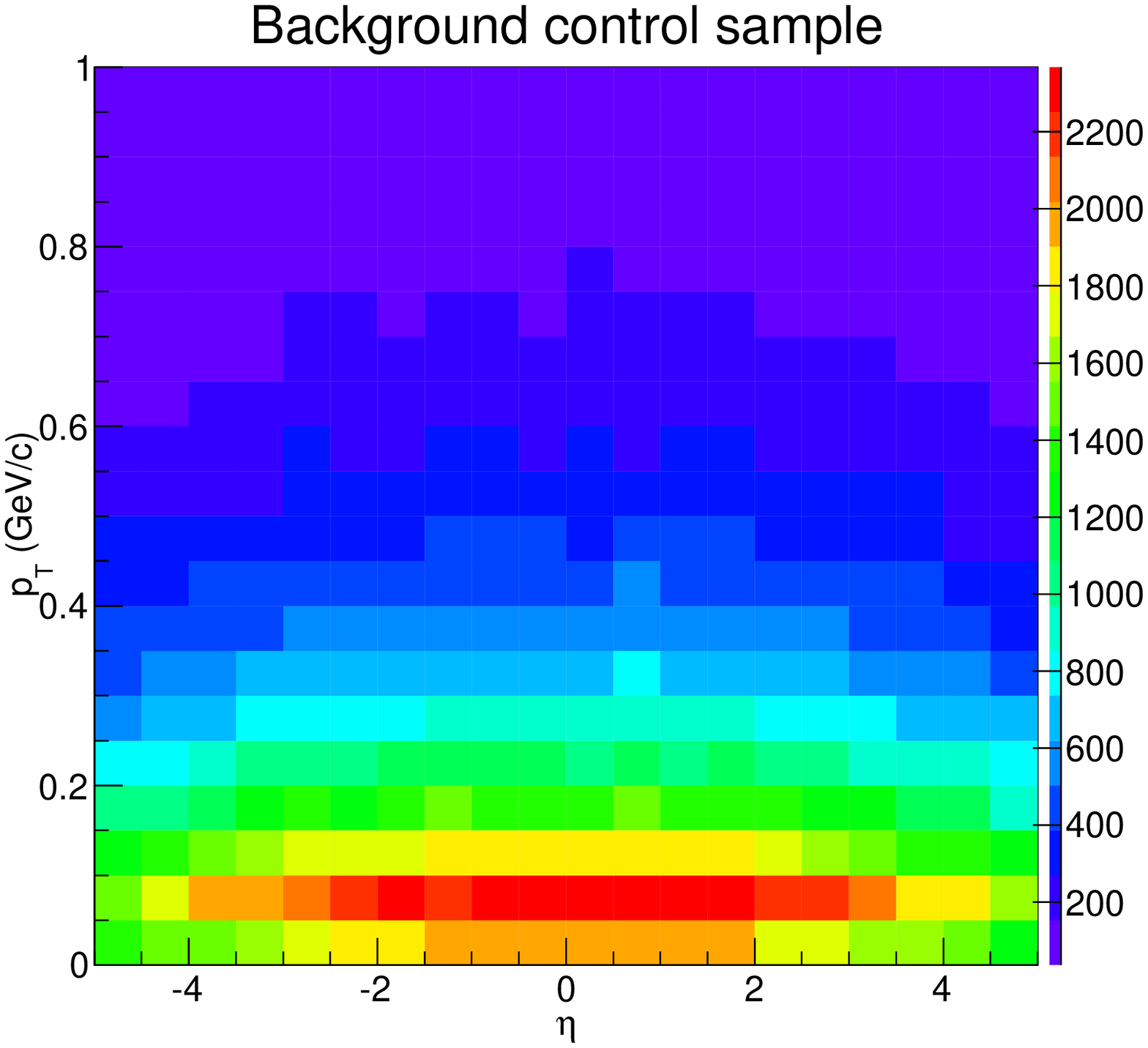}
\label{fig:cs:a}
}
\subfloat[]{
\includegraphics[scale=0.3]{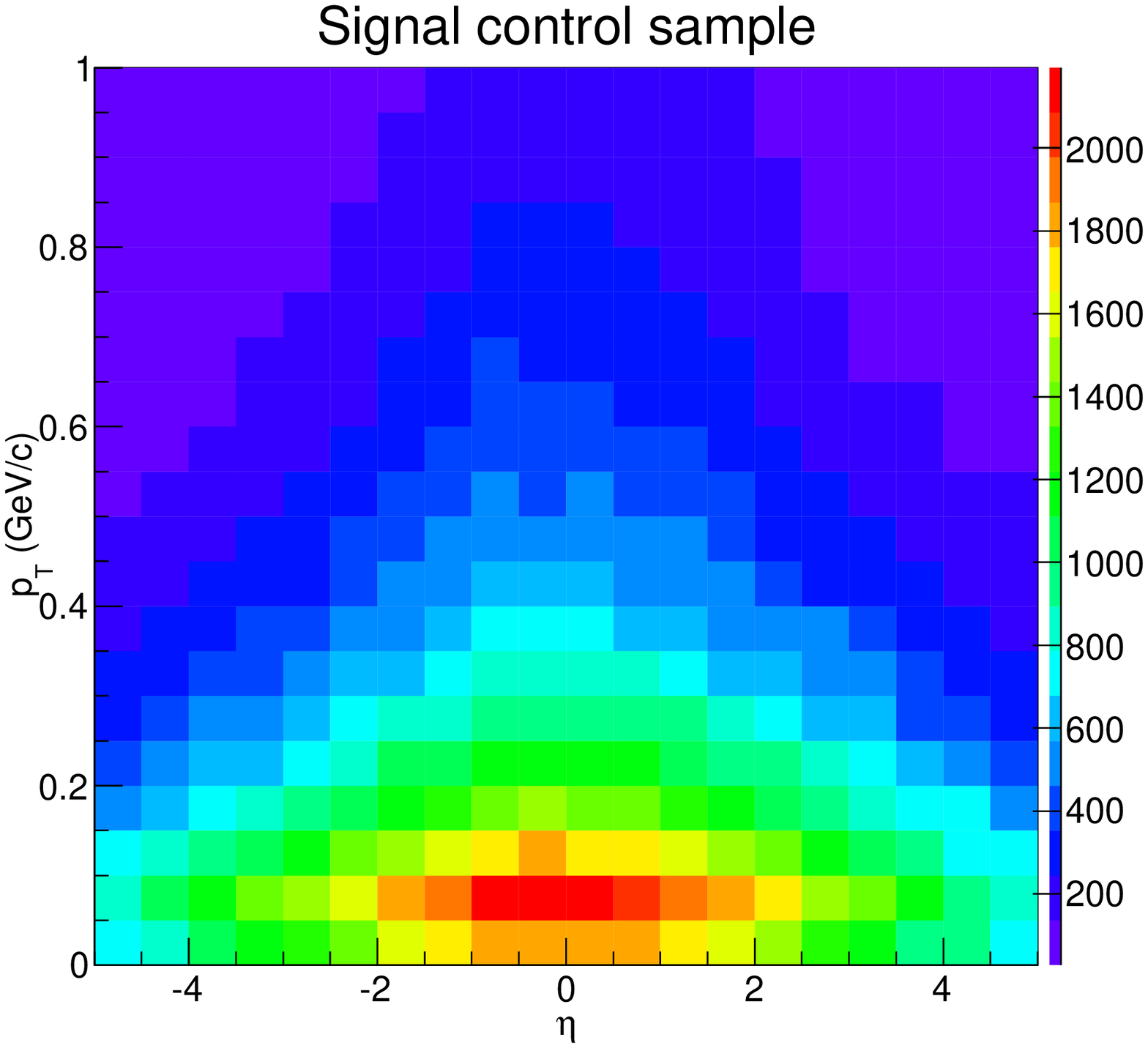}
\label{fig:cs:b}
}
\caption[]{
Particle-level $(\eta, p_T)$ distributions from the control samples in the region $-5<\eta<5$, $0<p_T<1~\mbox{GeV/c}$, 
as described in the text. (a) Neutral soft QCD particles. (b) Neutral particles from the hard parton scattering.
}
\label{fig:cs}
\end{figure*}

Signal and background PDF templates as functions of particle $\eta$ and $p_T$ were obtained using a control sample 
dataset containing ${\sim}300,000$ particles from $gg\rightarrow t\bar{t}$ and ${\sim}300,000$ from soft QCD interactions. 
The corresponding Monte Carlo truth ($\eta, p_T$) distributions of neutral soft QCD particles and of neutral particles 
from the hard scattering are shown in Fig. \ref{fig:cs}(a) and Fig. \ref{fig:cs}(b), respectively, each 
normalised to unit volume. 

The above-mentioned collections of ${\sim}300,000$ particles, although significantly-lower statistics than the large 
datasets normally used in the field, are considered adequate for the purpose of estimating the signal and background probability 
distributions in the context of this study. In fact, our emphasis is on how the soft QCD frequency distributions 
inside individual events deviate from the corresponding probability distribution. For this purpose, ${\sim}300,000$ particles 
are high-enough statistics for the local features in the frequency distributions due to the presence of statistical 
fluctuations to be averaged out.

\begin{figure*}
\centering
\subfloat[]{
\includegraphics[scale=0.3]{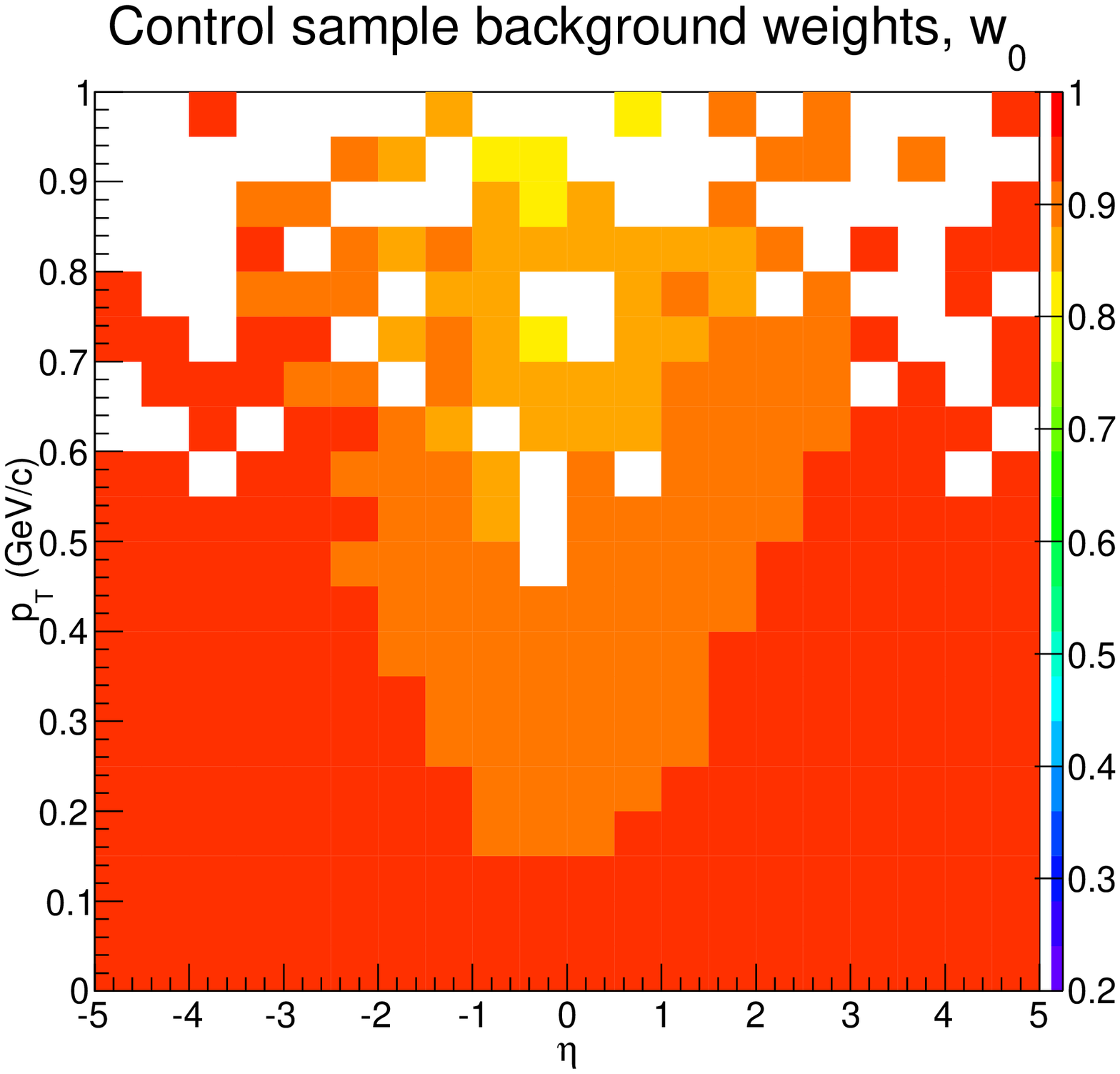}
}
\subfloat[]{
\includegraphics[scale=0.3]{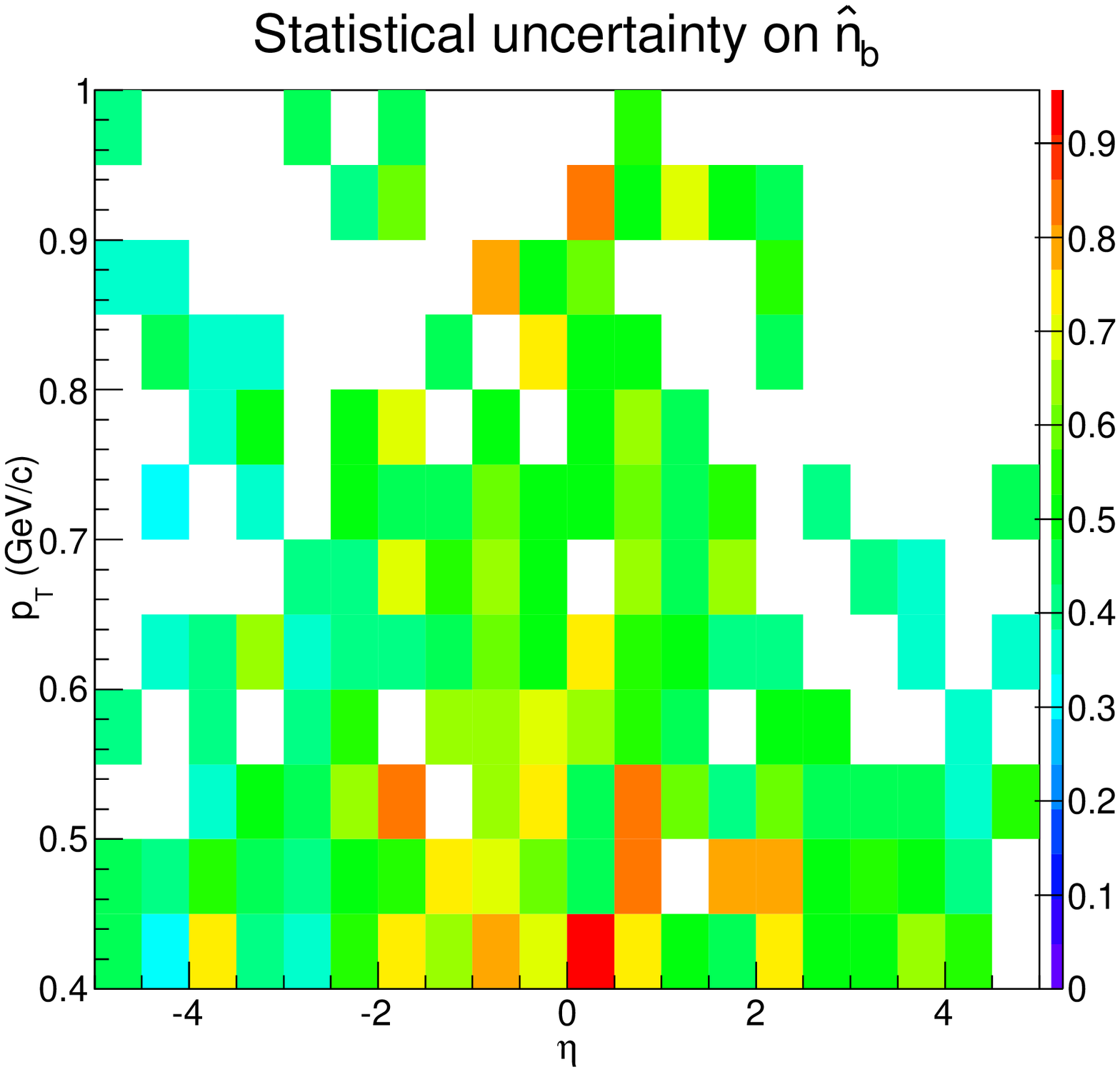}
}
\caption[]{
(a) Distribution of $w_0(\eta, p_T)$, estimated as described in the text. (b) Statistical uncertainty 
on the estimated number of soft QCD particles, $\sigma_{n_b}$, represented as a heat map across the particle 
$(\eta, p_T)$ space, inside the event chosen to illustrate our results. Additional information is provided in the text. 
}
\label{fig:heatmap}
\end{figure*}

The 
distributions 
shown in Fig. \ref{fig:cs} were used together with the previously-mentioned estimate of the average fraction of 
soft QCD particles over all neutrals in the events, $\alpha_0$, in order to calculate $w_0$ according to expression 
(\ref{eq:w0}). The distribution of $w_0(\eta, p_T)$ is shown in Fig. \ref{fig:heatmap}(a) in relation to the Monte Carlo 
event chosen to illustrate our results.

\begin{figure*}
\centering
\subfloat[]{
\includegraphics[scale=0.3]{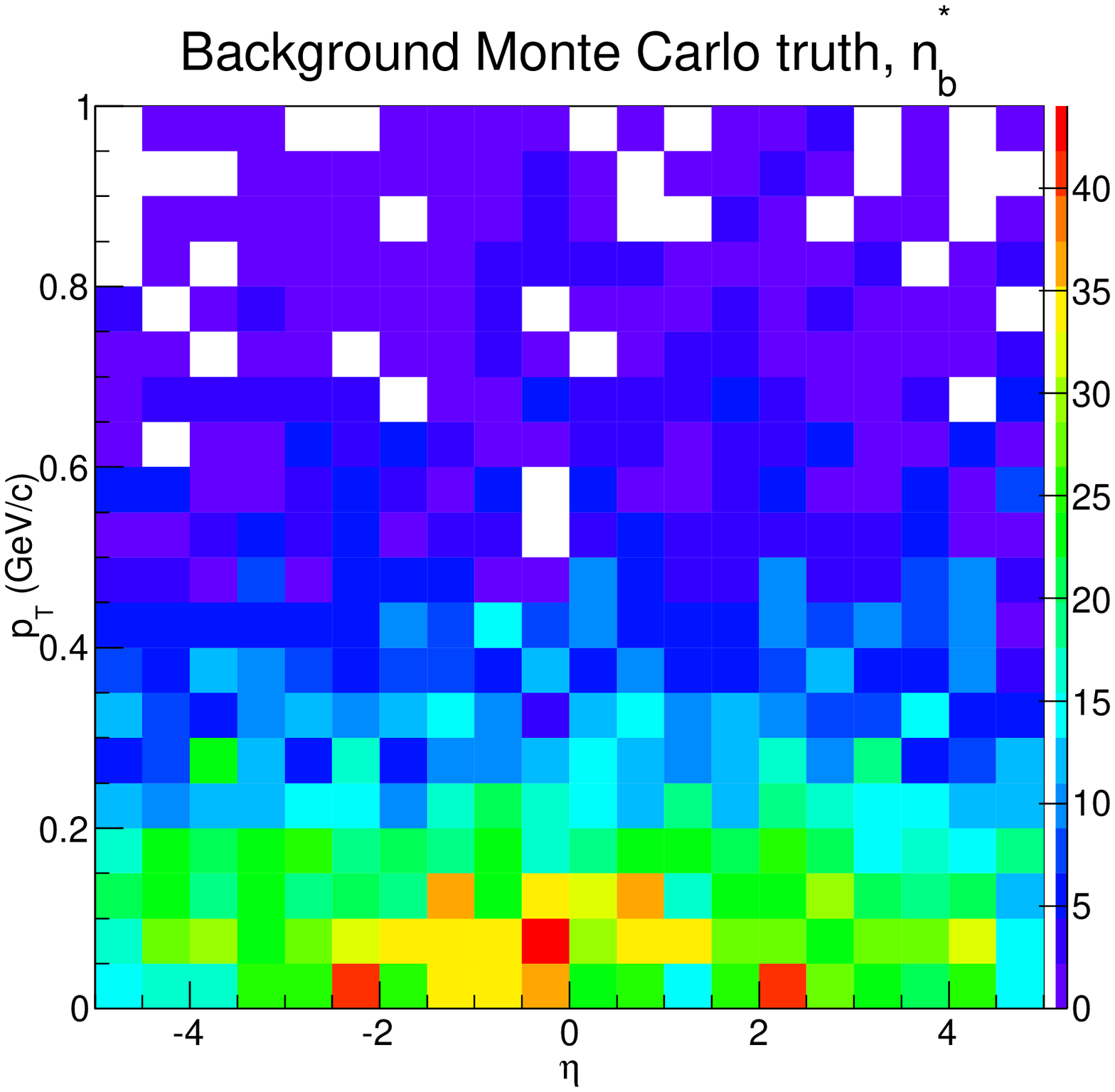}
}
\subfloat[]{
\includegraphics[scale=0.3]{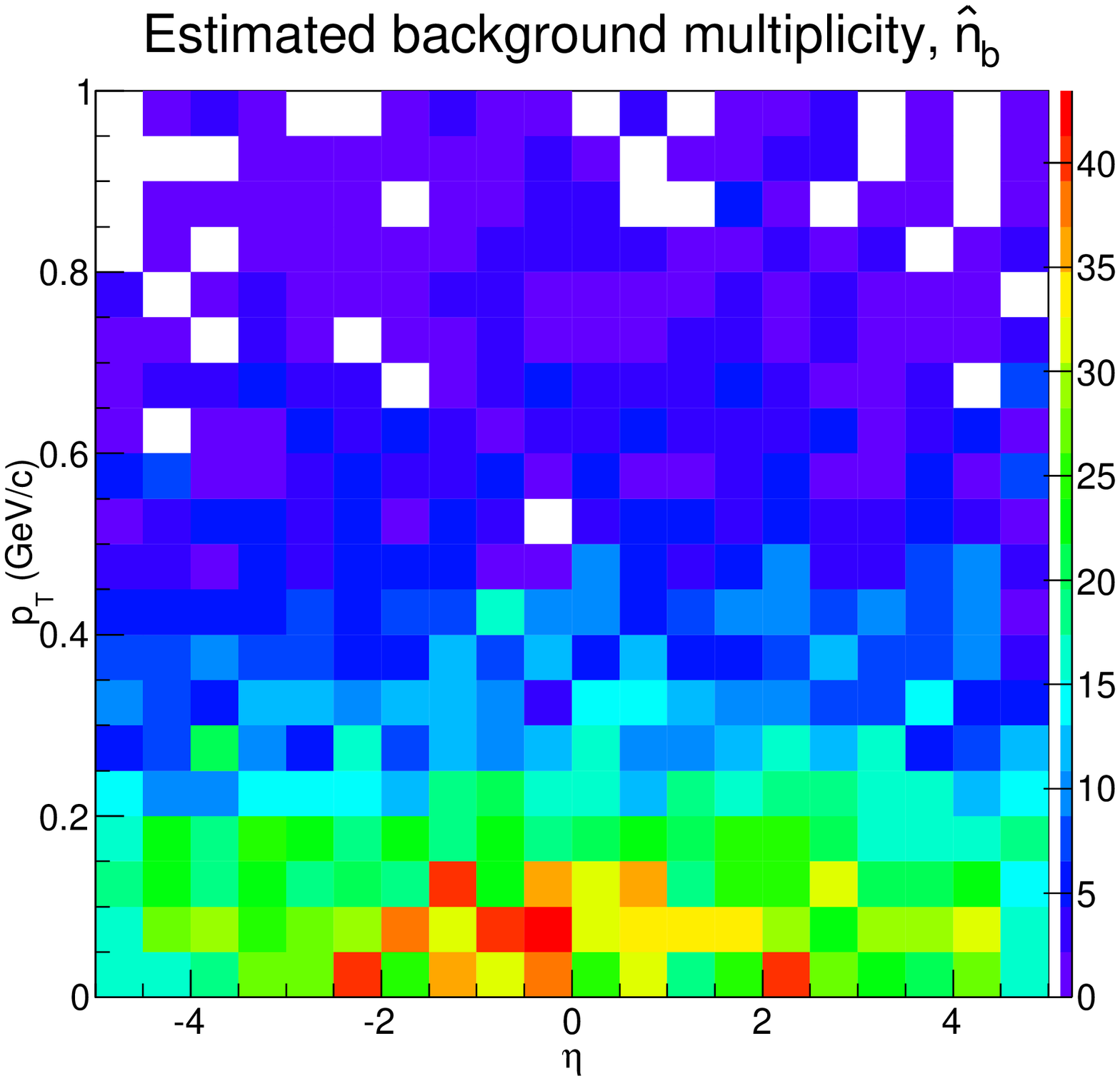}
}
\caption[]{
(a) True particle-level $(\eta, p_T)$ distribution of neutral soft QCD particles corresponding to one of the Monte Carlo 
events generated in this study. The plot illustrates how the soft QCD particle multiplicity 
across the kinematic space inside a typical event deviates from the corresponding 
probability distribution due to the limited particle statistics. 
(b) The corresponding particle-level $(\eta, p_T)$ distribution of neutral soft QCD particles estimated using this 
algorithm as described in the text.
}
\label{fig:truth}
\end{figure*}

\begin{figure*}
\centering
\subfloat[]{
\includegraphics[scale=0.3]{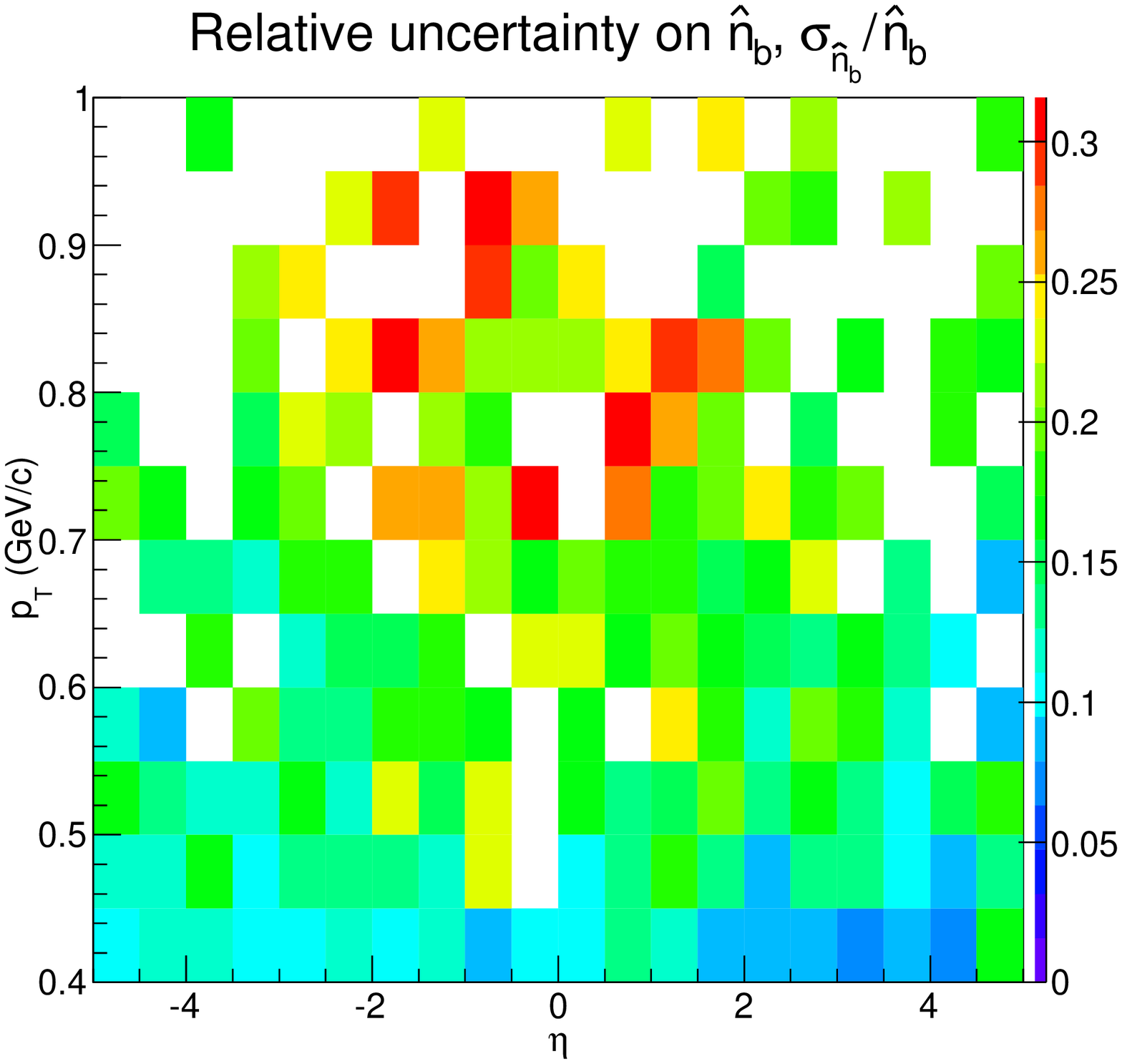}
}
\subfloat[]{
\includegraphics[scale=0.3]{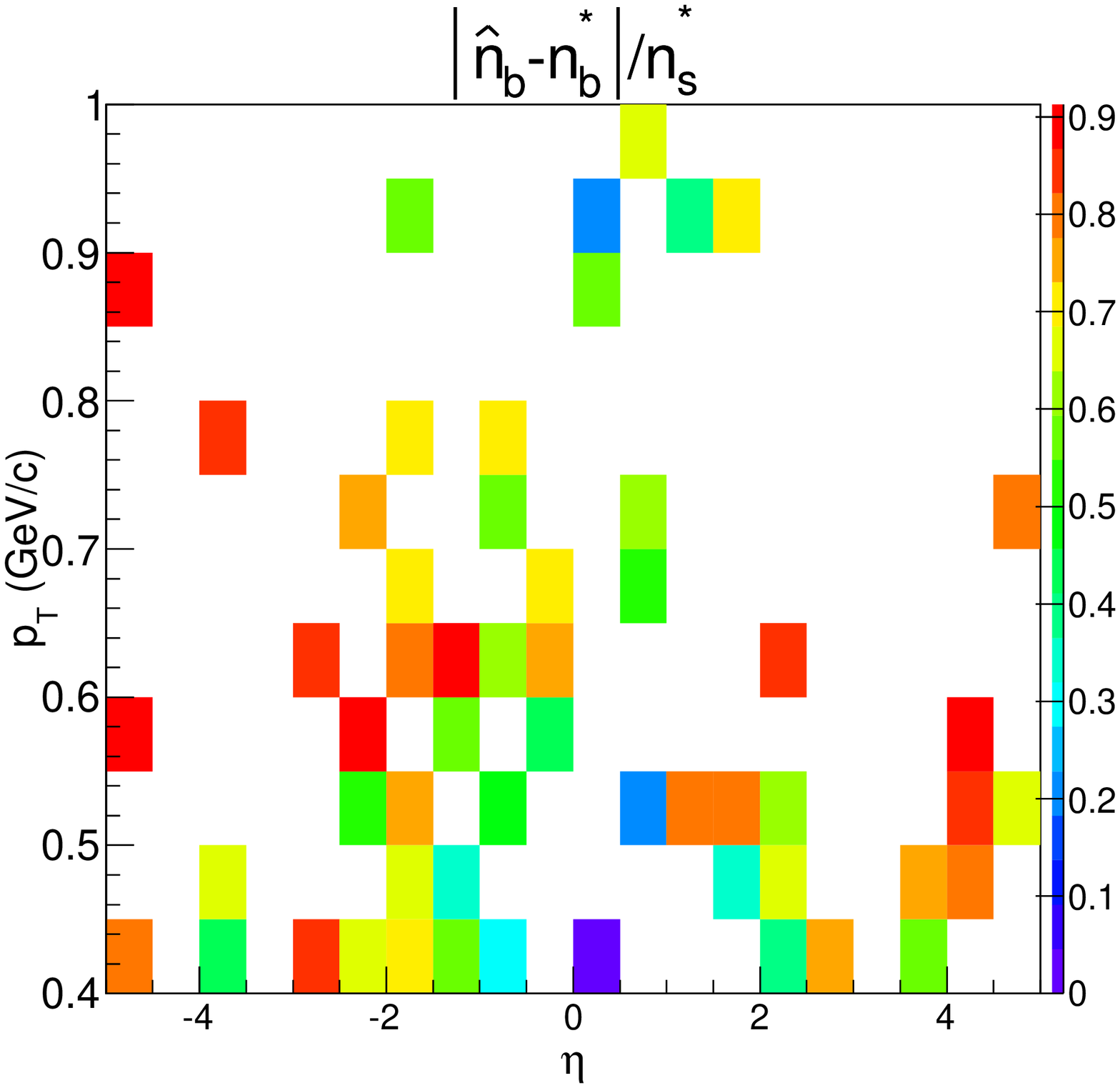}
}
\caption[]{
(a) Relative statistical uncertainty on the estimated number of soft QCD particles, $\sigma_{n_b}/\hat{n}_b$, 
represented as a heat map across the particle $(\eta, p_T)$ space, inside the event chosen to illustrate our results. 
(b) The corresponding absolute deviation of $\hat{n}_b$ from the corresponding true value, normalised to the true number 
of signal particles, $\left|\hat{n}_b-n^*_b\right|/n^*_s$, in those $(\eta, p_T)$ bins that contain at least 1 signal particle.
}
\label{fig:heatmap2}
\end{figure*}

Fig. \ref{fig:truth}(a) displays the true multiplicity of neutral soft QCD particles as a function of particle $\eta$ and $p_T$
in the reference event. 
As expected, the frequency distribution deviates from the corresponding 
higher-statistics distribution, 
shown in Fig. \ref{fig:cs}(a), 
due to the presence of local features that are typically washed out when multiple events are lumped together.

The estimate of the 
multiplicity of neutral soft QCD particles across the $(\eta, p_T)$ space 
obtained using this method is shown in Fig. \ref{fig:truth}(b)
with reference to the same event. 
A comparison with the corresponding true distribution in Fig. \ref{fig:truth}(a) suggests that the local features of 
the distribution are reasonably well described, e.g.\ the excess at $\eta\simeq 2.5$ and $p_T\simeq 0.2~\mbox{GeV}/c$. 
The performance is discussed further below with reference to all the events generated.

The absolute and relative statistical uncertainties on $\hat{n}_b$ in the reference event are displayed in 
Fig. \ref{fig:heatmap}(b) and Fig. \ref{fig:heatmap2}(a), respectively, where the absolute uncertainty, 
$\sigma_{n_b}$, was estimated using expression (\ref{eq:sigma_nbhat}). In particular, Fig. \ref{fig:heatmap}(b) 
suggests that the precision of this method is better than 1 particle, although, in order for this claim to be made, 
the precision over all events generated also needs to be assessed, as discussed in the following.

It is worth emphasising that, in general, it is not known which $(\eta, p_T)$ bins in the event contain signal particles 
and which do not. Therefore, once it is observed that $\hat{n}_b$ is a better estimate of $n^*_b$ than $\nu_b$ is, 
and that it is sufficiently precise, it is also necessary to verify that $\hat{n}_b$ is more accurate than the estimate 
that would be obtained if the possible presence of signal particles in the bins was simply neglected. 
For this purpose, the absolute deviation of the estimated number of neutral soft QCD particles from the true value, 
normalised to the true number of signal particles, $\left|\hat{n}_b-n^*_b\right|/n^*_s$, is displayed in 
Fig. \ref{fig:heatmap2}(b) above $0.4~\mbox{GeV}/c$ particle $p_T$ in those ($\eta, p_T$) bins that contain at least 
1 signal particle. As it can be seen, $\left|\hat{n}_b-n^*_b\right|/n^*_s \lesssim 1$, i.e.\ the absolute deviation 
of the estimated number of background particles from the true number is lower than the number of signal particles 
across the kinematic space in the event, corresponding to the pile-up rate considered.
\begin{figure*}
\centering
\subfloat[]{
\includegraphics[scale=0.3]{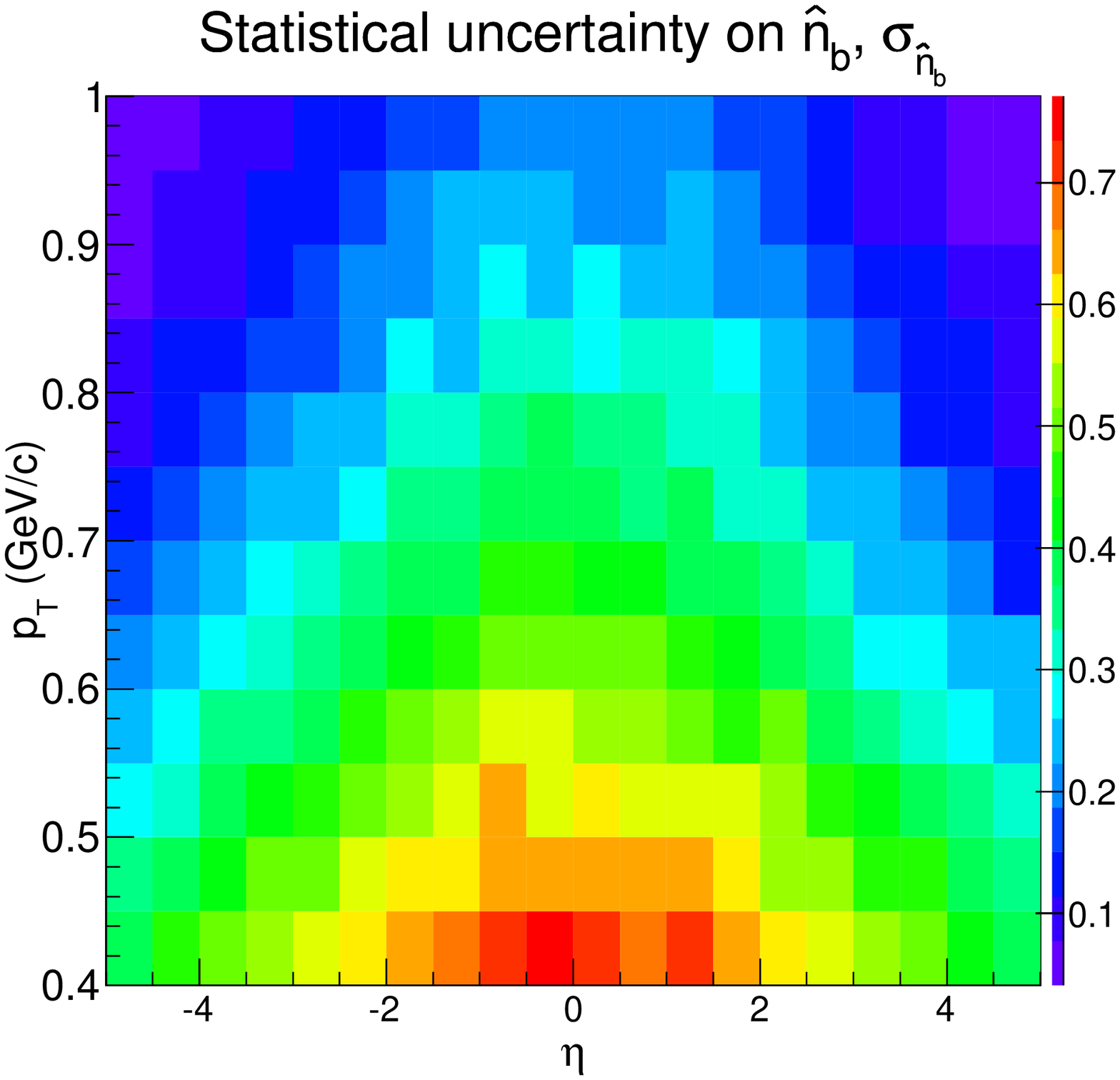}
}
\subfloat[]{
\includegraphics[scale=0.3]{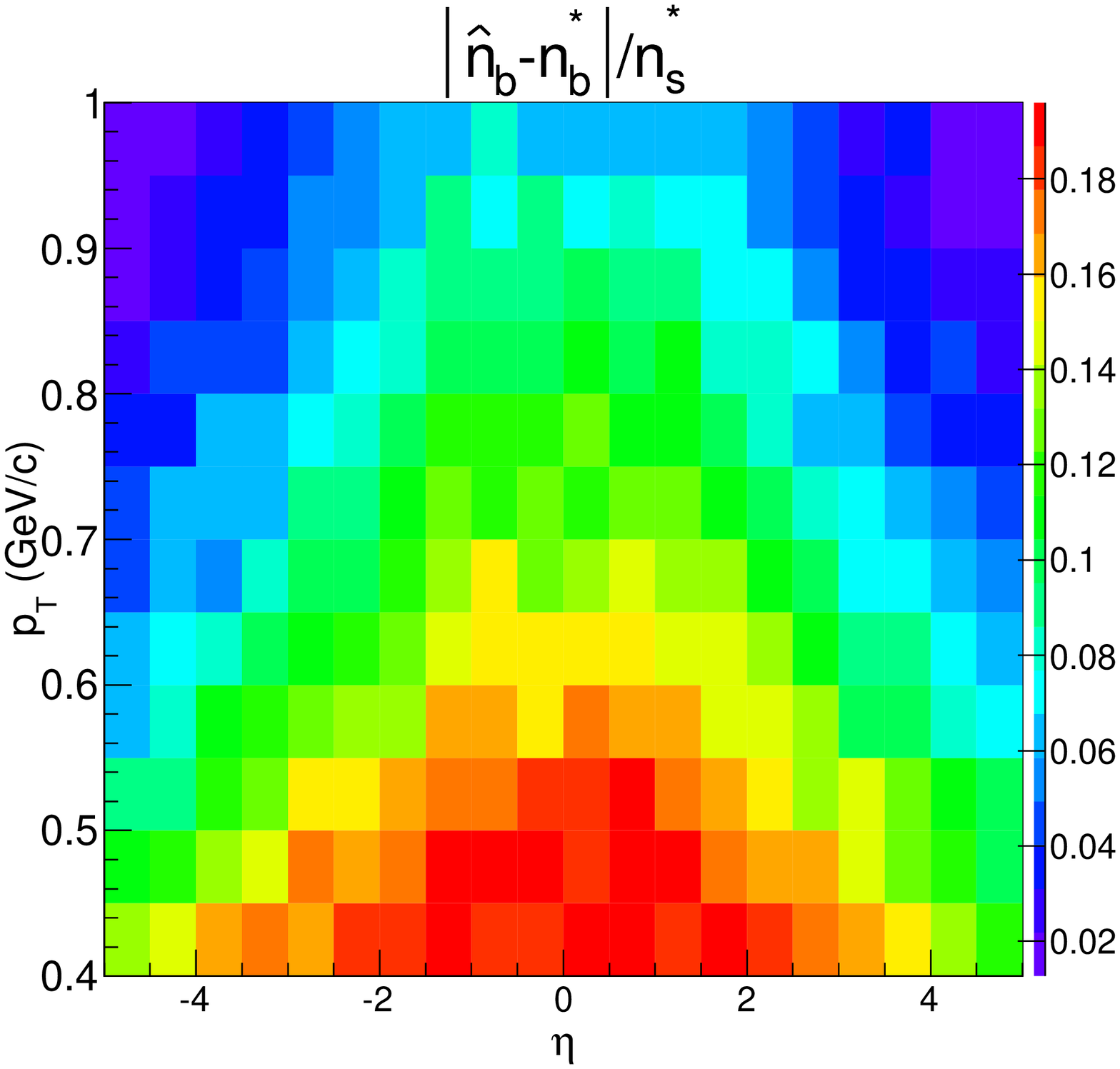}
}
\caption[]{
(a) Statistical uncertainty on the estimated number of soft QCD particles, $\sigma_{n_b}$, as a function of particle 
$\eta$ and $p_T$, averaged over the events generated for this study. (b) Absolute deviation of the estimated number of soft 
QCD particles from the corresponding true value, normalised to the true number of signal particles, 
$\left|\hat{n}_b-n^*_b\right|/n^*_s$, as a function of particle $\eta$ and $p_T$, averaged over the events generated 
for this study. The plots confirm the level of precision and accuracy that can be achieved using this method, 
in that $\sigma_{n_b}$ is below 1 particle and $\left|\hat{n}_b-n^*_b\right|/n^*_s$ is significantly lower than 1.
}
\label{fig:precision}
\end{figure*}
As anticipated, Fig. \ref{fig:heatmap}(b) and Fig. \ref{fig:heatmap2} relate to one single event. In order to verify 
the precision and the accuracy of the algorithm in more detail, the corresponding heat maps were obtained over all 
events generated. Fig. \ref{fig:precision}(a) and Fig. \ref{fig:precision}(b) display 
$\left<\sigma_{n_b}\right>$ and $\left<\left|\hat{n}_b-n^*_b\right|/n^*_s\right>$ across the $(\eta, p_T)$ kinematic space, 
respectively, where the average is taken over events. The plots confirm that the algorithm produces consistent results, 
with $\left<\sigma_{n_b}\right>$ below 1 particle and $\left<\left|\hat{n}_b-n^*_b\right|/n^*_s\right>$ significantly lower 
than 1 across the ($\eta, p_T$) space. 

It is worth noticing that the present investigation relies on a particle-level kinematic comparison between soft QCD interactions and a specific hard scattering process, namely $gg\rightarrow t\bar{t}$. Choosing a different signal process will normally change the final-state particle kinematics, although the difference between particles originating from a hard scattering and soft QCD particles is generally expected to be more pronounced than differences across signal processes. As discussed in section \ref{algo}, this is supported by the study documented in \cite{CHEP2015_arxiv}, where this method was applied to vector boson fusion Standard Model Higgs production. In any case, the potential dependence of the performance of this technique on the choice of signal process deserves further investigation, in order for the results presented in this article to be generalised.


\section{Conclusion\label{concl}}

The contamination, or background, from particles produced by low-energy strong interactions is a major issue at the 
Large Hadron Collider. Although well-established correction procedures are in use at the experiments, new techniques are 
also being investigated in the field. The primary objective of this new line of development is to meet the requirements 
that will be posed by the upcoming higher-luminosity operational regimes of the accelerator.

We have investigated a different perspective to mainstream methods, thereby estimating the kinematics of background particles 
inside individual collision events in terms of their frequency distribution. Whereas the use of probability distributions 
is traditional, our emphasis on the frequency distributions is, to the best of our knowledge, a distinctive 
and unique feature of our approach.

Our hope and expectation is that the ability to describe the kinematic frequency distribution of the contaminating particles 
collision by collision will help towards the development of improved subtraction methods at higher luminosity. 
In particular, we expect that our method will become useful as a complement to existing correction algorithms applied to jets, 
i.e.\ to collections of final-state particles interpreted as originating from the same scattered parton.

The preliminary results discussed in this article suggest that our method can produce more accurate estimates of the 
number of contaminating particles in different kinematic regions inside collision events than would be possible by relying 
exclusively on the expected numbers. Although the possible impact of mismodelling remains to be investigated in more detail, 
this encourages further studies in this direction. 

However, it should be stressed that a proper quantitative test of the proposed method will require a detailed assessment on specific observables, which is beyond the scope of the present feasibility study.

It should also be emphasised that the algorithm is inherently parallel. In fact, different kinematic regions inside events 
can be processed independently, and the calculation of the only global variable, i.e.\ of the average fraction of 
contaminating particles per event, can be performed in advance on a control sample. For this reason, this method is potentially 
suitable for inclusion in future particle-level event filtering procedures upstream of jet reconstruction.

The possible complementarity between our method and the particle weighting techniques that have recently been proposed at 
the Large Hadron Collider will also be worth exploring. In fact, our estimate of the probability for individual particles 
to originate from low-energy QCD processes as opposed to the high-energy signal scattering is based on information that is 
currently not employed by any particle weighting algorithms. In this context, we anticipate that multivariate combinations 
of different weighting schemes can prove beneficial with a view to improving further on the rejection of contaminating particles.

Finally, it will be useful to study the impact of this method, when used in conjunction with existing techniques, 
on the resolution of the missing transverse energy as well as on estimates of particle isolation, with a view to 
quantifying the associated benefit at the level of physics analysis. The source code used for the purpose of this study is made available upon request.

\section{Competing interests}
The authors declare that there is no conflict of interest regarding the publication of this paper.


\begin{thebibliography}{1}
\bibitem{REVIEW_pile-up_2014}The CMS Collaboration, ``Pileup Removal Algorithms'', PAS JME-14-001, 2014
\bibitem{jet-area}M.~Cacciari and G.~P.~Salam, ``Pileup subtraction using jet areas'', {\it Phys. Lett.} B, vol. 659, no. 1-2, pp. 119-126, 2008
\bibitem{jet-filtering}J.~M.~Butterworth, A.~R.~Davison, M.~Rubin and G.~P.~Salam, ``Jet substructure as a new Higgs search channel at the LHC'', {\it Phys. Rev. Lett.}, vol. 100, no. 24, 2001, 2008
\bibitem{jet-trimming}D.~Krohn, J.~Thaler and L.-T.~Wang, ``Jet Trimming'', {\it J. High Energy Phys.}, vol. 2010, no. 2, 84, 2010
\bibitem{jet-pruning-1}S.~D.~Ellis, C.~K.~Vermilion and J.~R.~Walsh, ``Techniques for improved heavy particle searches with jet substructure'', {\it Phys. Rev.} D, vol. 80, no. 5, 1501, 2009
\bibitem{jet-pruning-2}S.~D.~Ellis, C.~K.~Vermilion and J.~R.~Walsh, ``Recombination Algorithms and Jet Substructure: Pruning as a Tool for Heavy Particle Searches'', {\it Phys. Rev.} D, vol. 81, no. 9, 4023, 2010
\bibitem{jet-soft-drop}A.~J.~Larkoski, S.~Marzani, G.~Soyez and J.~Thaler, ``Soft Drop'', {\it J. High Energy Phys.}, vol. 2014, no. 5, 146, 2014
\bibitem{jet-cleansing}D.~Krohn, M.~D.~Schwartz, M.~Low and L.-T.~Wang, ``Jet cleansing: Separating data from secondary collision induced radiation at high luminosity'', {\it Phys. Rev.} D, vol. 90, no. 6, 5020, 2014
\bibitem{PUPPI}D.~Bertolini, P.~Harris, M.~Low and N.~Tran, ``Pileup per particle identification'', {\it J. High Energy Phys.}, vol. 2014, no. 10, 59, 2014
\bibitem{SoftKiller}M.~Cacciari, G.~P.~Salam and G.~Soyez, ``SoftKiller, a particle-level pileup removal method'', {\it Eur. Phys. J. C}, vol. 75, 59, 2015 
\bibitem{berta}P.~Berta, M.~Spousta, D.~W.~Miller and R.~Leitner, ``Particle-level pileup subtraction for jets and jet shapes'', {\it J. High Energy Phys.}, vol. 2014, no. 6, 92, 2014 
\bibitem{pythia1}T.~Sjöstrand, S.~Mrenna and P.~Skands, ``PYTHIA 6.4 physics and manual'', {\it J. High Energy Phys.}, vol. 2006, no. 5, 26, 2006
\bibitem{pythia2}T.~Sjöstrand, S.~Mrenna and P.~Skands, ``A brief introduction to PYTHIA 8.1'', {\it Comput. Phys. Comm.}, vol. 178, no. 11, pp. 852-867, 2008
\bibitem{gibbshep3}F.~Colecchia, ``Toward particle-level filtering of individual collision events at the Large Hadron Collider and beyond'',  {\it J. Phys.: Conf. Ser.}, vol. 490, 012226, 2014
\bibitem{gibbshep2}F.~Colecchia, ``A sampling algorithm to estimate the effect of fluctuations in particle physics data'', {\it J. Phys.: Conf. Ser.}, vol. 410, 012028, 2013
\bibitem{gibbshep}F.~Colecchia, ``A population-based approach to background discrimination in particle physics'', {\it J. Phys.: Conf. Ser.}, vol. 368, 012031, 2012
\bibitem{CHEP2015_arxiv}F.~Colecchia, ``Data-driven estimation of neutral pileup particle multiplicity in high-luminosity hadron collider environments'', {\it J. Phys.: Conf. Ser.}, vol. 664, 072013, 2015
\bibitem{fingerprints_arxiv}F.~Colecchia, ``Particle-level kinematic fingerprints and the multiplicity of neutral particles from low-energy strong interactions'', ({Preprint} arXiv:1412.1989 [hep-ph]), 2015
\end{thebibliography}
\end{document}